# Spin-glass-like behaviour in SmFeAsO$_{0.8}$F$_{0.2}$

**Nikolai D. Zhigadlo*,[a] and Roman Puzniak[b]**

[a] *CrystMat Company, CH-8037 Zurich, Switzerland*

[b] *Institute of Physics, Polish Academy of Sciences, PL-02668 Warsaw, Poland*

**Graphical Abstract**

The iron-based oxypnictide superconductor SmFeAsO$_{0.8}$F$_{0.2}$ was synthesized under high-pressure and investigated by measuring *dc* magnetic susceptibility. The zero-field cooled (ZFC) magnetic susceptibility confirmed the bulk superconductivity of the sample with a critical temperature of $T_c \simeq 50$ K and a significant jump in the magnetization at ~4.3 K, usually attributed to the antiferromagnetic ordering of Sm$^{3+}$ ions in this system. Since the occurrence of the jump depends on the cooling history, our data strongly suggest a spin-glass-like behaviour.

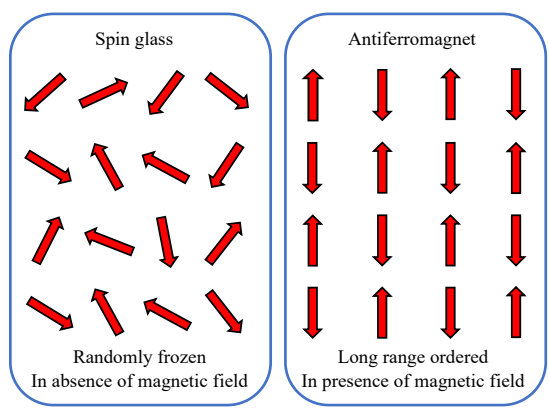

* nzhigadlo@gmail.com; http://crystmat.com

For many decades, the study of spin glasses has been an attractive topic in magnetism.[1-5] Basically, a spin glass can be defined as a set of interacting magnetic moments originating from spins, in which the interactions are randomly distributed in sign and possibly in magnitude.[5] The many available spin arrangements with comparable energies produce a huge number of metastable states. Hence, finding the absolute minimum is extremely difficult, and a spin glass is always out of equilibrium. The first spin glass materials discovered were nonmagnetic metals (Au, As, Pt, *etc*.), in which a few percent of magnetic atoms (Fe, Mn, *etc*.) were randomly dispersed.[6] At present, many different systems are known to exhibit spin-glass or spin-glass-like behaviour; among them are metals including superconductors, metallic glasses, dilute magnetic alloys, semimetals and insulating glasses. All these materials exhibit particular dynamics of magnetization, and many interesting effects are observed in magnetic susceptibility.[5]

The discovery of Fe-based high-temperature superconducting materials has opened a new window into the physics of unconventional superconductors. Many of these materials contain at least two magnetic elements (Fe and a magnetic rare-earth metal). Spin-glass behaviour due to spin disorder and magnetic frustration below a certain freezing temperature has been reported in Fe-based superconductors such as P-doped $EuFe_2As_2$,[7] Co-doped $BaFe_2As_2$,[8] Se-doped FeTe[9] and Ce-doped $CaFe_2As_2$.[10] The parent compound SmFeAsO is one of the members of a large family of *Ln*FeAsO (*Ln*1111, *Ln*: lanthanide) exhibiting two types of magnetic orders: antiferromagnetic (AFM) ordering of the Fe spins in the *ab*-plane at ~130 K and ordering of Sm spins along the *c*-axis at ~5 K (Figure 1).[11,12]

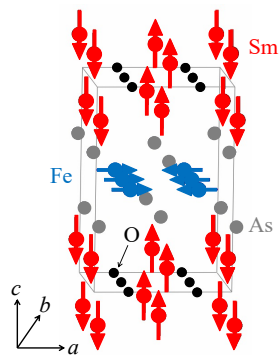

**Figure 1** Magnetic structures adopted by the Fe and Sm sublattices in SmFeAsO below 5 K (two unit cells are shown).

The AFM ordering of Fe spins can be suppressed upon appropriate doping; hence, superconductivity emerges at the expense of antiferromagnetism, whereas Sm ordering survives.[13,14] Interestingly, Nd moments also align in the *c* direction in NdFeAsO compounds with the same AFM alignment as in SmFeAsO compounds, *i.e.* low-temperature magnetism in these compounds may share a similar underlying physics.[15] Despite extensive experimental efforts over the past years, the interplay between superconductivity and magnetism in the *Ln*FeAsO family remains an important open issue, which requires further exploration.

When studying the SmFeAsO$_{0.8}$F$_{0.2}$ superconductor by measuring the standard *dc* magnetic susceptibility using both zero-filed-cooled (ZFC) and field-cooled (FC) protocols, we observed a significant jump in magnetization at ~ 4.3 K. Such a jump was interpreted as a result of the antiferromagnetic ordering of Sm$^{3+}$ ions in this system,[16] which mimics electron-doped high-$T_c$ cuprate Sm$_{2-x}$Ce$_x$CuO$_{4-\delta}$.[17] In our opinion, the situation is somewhat more complicated if we take into account that the appearance of the magnetization jump depends on the cooling history of the sample. The present data suggest a spin-glass-like behaviour of SmFeAsO$_{0.8}$F$_{0.2}$ in the low temperature region.

A polycrystalline sample with the empirical formula SmFeAsO$_{0.8}$F$_{0.2}$ was prepared in a cubic anvil high-pressure cell. The sample preparation and the packing of a high-pressure cell assembly were performed in a glove box with a protective argon atmosphere. A stoichiometric mixture of high purity (99.95%) SmAs, FeAs, Fe$_2$O$_3$, Fe and SmF$_3$ powders was loaded into a boron nitride (BN) crucible and placed in a pyrophillite sample cube with a graphite furnace, which was pressurized to 3 GPa. The crucible was heated to 1350 °C in 1 h and kept at this condition for 4.5 h; then it was quenched to room temperature. After completing the synthesis, the pressure was released and the sample was removed. X-ray measurements revealed high homogeneity and the single-phase crystalline nature of the sample, and the absence of significant amounts of impurities. The resulting sample stoichiometry was revealed by energy-dispersive X-ray spectroscopy (EDX) analysis and further confirmed by X-ray structure refinement (for details, see Online Supplementary Materials). Further details about sample preparation and the experimental setup can be also found in our previous publications.[18-20] The temperature dependence of the magnetic susceptibility of a powdered polycrystalline SmFeAsO$_{0.8}$F$_{0.2}$ sample of 116 mg was measured on a Quantum Design Physical Property Measurement System (PPMS) equipped with Reciprocating Sample Option (RSO) in an external magnetic field of 10 Oe using both ZFC and FC protocols. To eliminate the possible influence of non-uniformities of the device magnet on the measurement results, especially at very low fields, efforts were made to minimize the remnant field.

Figure 2(a) shows the characteristic change in magnetization as a function of temperature. The sample was cooled in a zero field from the temperature region of the paramagnetic state to a temperature of 1.8 K. After reaching the desired temperature and waiting for 1 h, the evolution of the magnetization was measured under a tiny applied magnetic field of 10 Oe. The measurement reveals the bulk nature of superconductivity with a critical temperature $T_c \simeq 50$ K. The reduced low-temperature diamagnetic response, equal to ~ 27% of the ideal superconductivity response, is due to the relatively small grain size of the material under study, comparable to its penetration depth. The low-temperature part of the ZFC curve is shown in Figure 2(b), where a small but distinct jump in magnetization is observed at ~ 4.3 K (curve 1).

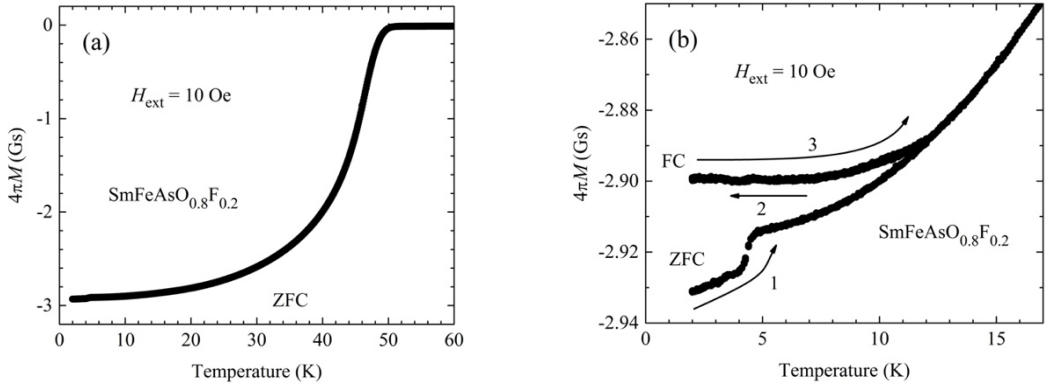

**Figure 2** Temperature dependence of the magnetization of SmFeAsO$_{0.8}$F$_{0.2}$ recorded using the standard ZFC and FC protocols. (a) The sample was first cooled in a zero field to 1.8 K, and measurements were carried out while heating it after applying an external field of 10 Oe. (b) Measurements were performed in a magnetic field of 10 Oe (1) upon heating after ZFC, (2) then upon cooling in the field and (3) further upon reheating in the field. A small but distinct jump in magnetization occurs at ~ 4.3 K. The onset of irreversibility occurs at ~ 12 K, where the ZFC-FC curves diverge.

In the published data, this transition is usually assigned to the AFM ordering of the Sm$^{3+}$ ions in this system.[16,21] However, the present data suggest rather a spin-glass-like behaviour, since the appearance of the jump depends on the cooling history of the sample. It turned out that the magnetization jump can be removed by heating the SmFeAsO$_{0.8}$F$_{0.2}$ sample above the jump temperature and cooling it again. This is illustrated in Figure 2(b), which shows the results of measurements carried out in a magnetic field of 10 Oe upon heating after zero-field cooling (curve 1), then upon cooling in a field (curve 2), and further upon reheating in a field (curve 3). This behaviour suggests a large degree of disorder of the $Sm^{3+}$ magnetic moments and confirms the coexistence of spin-glass-like ordering and superconductivity in $SmFeAsO_{0.8}F_{0.2}$. At high temperatures, the Sm subsystem is in the paramagnetic phase and all the spins fluctuate. In our case, the ZFC curve [Figure 2 (b), curve 1] indicates nonequilibrium state the frozen phase, since the field was applied at low temperature. When the sample is cooled to low temperatures in the absence of an external field, many competing metastable states appear, separated by high-energy barriers, and the system can be tracked in any of them. By contrast, if the material is cooled in a non-zero external field, *i.e.*, in the FC mode [Figure 2(b), curve 2], the magnetization can be considered as equilibrium in the first approximation, and the $Sm^{3+}$ spins align with the magnetic field, which leads to the antiferromagnetic order.

In our case, the spin-glass-like feature is rather weak, and its possible origin can be ascribed to (i) random magnetic exchange interactions between inhomogeneously distributed $Sm^{3+}$ spins, (ii) magnetic exchange coupling between Fe and Sm spins, which leads to frustration among the Sm spins, or to both. The later case was actually observed in parent single crystalline SmFeAsO.[11] An X-ray resonant magnetic scattering experiment showed that due to the interconnection between Fe and Sm, the induced Sm moments or the coupled moments of Fe and Sm appear at a temperature much higher than the magnetic ordering temperature of Sm, $T_{Sm} \simeq 4.3$ K. A similar interplay between two magnetic sublattices was also observed in the NdFeAsO system.[22] Such coupling, however, can be ruled out for $SmFeAsO_{0.8}F_{0.2}$, since the F doping suppresses AFM ordering, and yet, below $T \leq 12$ K we can see that the ZFC and FC curves diverge [Figure 2(b)]. Thus, we can hypothesize that, in addition to the randomly distributed spins of Sm, the coupled moments of Fe and Sm can also contribute to the spin-glass-like behaviour of $SmFeAsO_{0.8}F_{0.2}$.

It is interesting to note that the magnetic structure of the Sm subsystem in SmFeAsO and SmFeAs(O,F) is essentially the same as that adopted by the Sm moments in $Sm_2CuO_4$,[17] and that all compounds exhibit an unusual insensitivity of $T_{Sm}$ to an externally applied magnetic field.[21,23] It is still unclear why and how the Sm ordering coexists with superconductivity in $SmFeAsO_{0.8}F_{0.2}$ and what role it plays in setting the high superconducting transition temperatures observed in this series.

In summary, the observed dependence of magnetization at low temperature on history provides evidence for the existence of a spin-glass-like state in superconducting $SmFeAsO_{0.8}F_{0.2}$. This system exhibits spin-glass-like features when the sample is cooled in zero-field conditions. The origin of this behaviour is probably due to the inhomogeneous distribution of $Sm^{3+}$ spins in the Sm sublattice, which is accompanied by weak interactions between the Fe and Sm spins in the Fe and Sm sublattices. Further studies are needed to identify the precise magnetic structure and understand the complex interplay between the Sm and Fe sublattices. Determining its nature is an important attempt to understand the magnetism and superconductivity of the *Ln*FeAsO family.

The authors would like to acknowledge support from the Laboratory for Solid State Physics, ETH Zurich, where this study was initiated, and thank T. Shiroka for reading the manuscript critically and for helpful comments.

**Electronic supplementary materials** *Mendeleev Commun*., 2022, **32**, 305-307

**Structure characterization by X-ray powder diffraction**

Powder X-ray diffraction (XRD) analysis was performed at room temperature on a STOE StadiP diffractometer (CuK$_\alpha$ radiation, $\lambda$ = 1.54056 Å). The synthesized sample was thoroughly ground in a mortar, and then the XRD pattern of the powder was measured. All peaks in the XRD pattern can be indexed based on a tetragonal unit cell (*P*4/*nmm*) of the ZrCuSiAs-type structure.

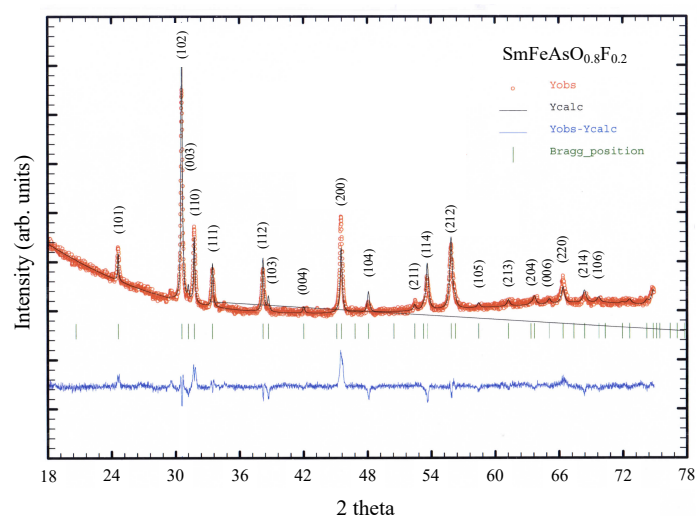

**Figure S1** Observed (open circles) and calculated (line) X-ray diffraction pattern of SmFeAsO$_{0.8}$F$_{0.2}$ sample with the difference at the same scale plotted below. The vertical bars indicate the angular positions of the allowed Bragg reflections. The profile analysis was carried out to refine the unit cell parameters: $a = b$ = 3.931(1) Å and $c$ = 8.468(1) Å.

**SEM and EDX characterization**

The morphology of SmFeAsO$_{0.8}$F$_{0.2}$ sample and the elemental distribution within the bulk material were investigated with the Hitachi S-3000 N scanning electron microscope (SEM) equipped with a Noran SIX NSS 200 dispersive X-ray detector. Cracking the bulk sample into small parts leads to the microstructure as illustrated in Figure 2S. It is composed of stacked randomly oriented *ab*-planes of grains with dimensions up to 50 × 50 × 15 μm$^3$.

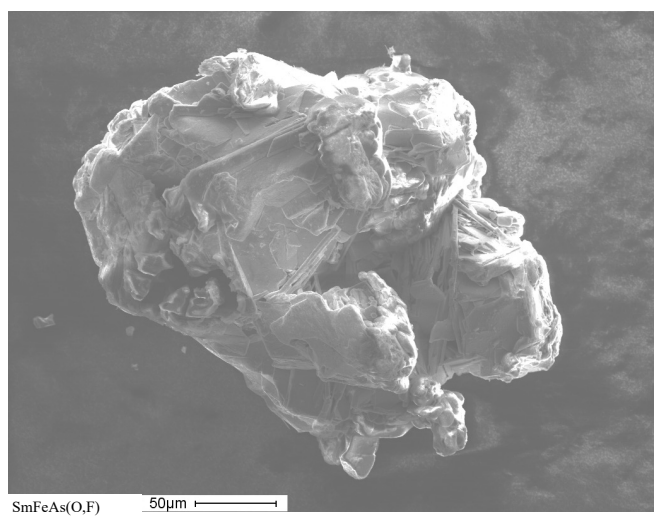

**Figure S2** Electron SEM image of SmFeAsO$_{0.8}$F$_{0.2}$ sample showing the typical microstructure.

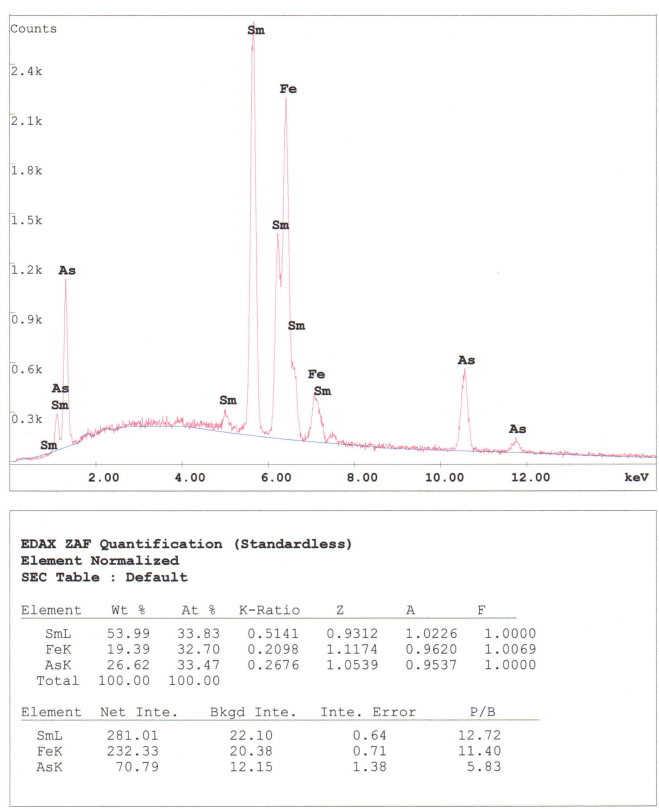

**Figure S3** Energy dispersive X-ray (EDX) analysis of SmFeAsO$_{0.8}$F$_{0.2}$ sample.

The resulting stoichiometry of the sample was revealed by energy dispersive X-ray spectroscopy (EDX) analysis and was further confirmed by X-ray structure refinement. Both methods show that the ratio of samarium, iron, and arsenic is equal 1:1:1, consistent with the nominal composition. The light elements of oxygen and fluorine cannot be detected accurately; therefore, for their content in the chemical formulae we quote the nominal values.